\documentclass[iop]{emulateapj}

\usepackage[utf8]{inputenc}
\usepackage{graphicx}
\usepackage{amsmath}
\usepackage{mathrsfs}
\usepackage{natbib}
\usepackage{array}
\usepackage{hyperref}
\usepackage{wasysym}

\usepackage{xcolor}

\def\imagetop#1{\vtop{\null\hbox{#1}}}

\slugcomment{First version}

\shorttitle{Can chondrules be produced by the interaction of Jupiter with the protosolar disk?}
\shortauthors{J.-D. Bod\'{e}nan et al.}

\begin{document}

\title{Can chondrules be produced by the interaction of Jupiter with the protosolar disk?}

\author{Jean-David Bod\'{e}nan}
\affil{Institut für Geochemie und Petrologie, Departement Erdwissenschaften, Clausiusstrasse 25, 8092, Zürich, Switzerland}
\affil{Institute for Computational Science, University of Zürich, Winterthurerstrasse 190, 8057 Zürich, Switzerland.}
\email{jean-david.bodenan@erdw.ethz.ch}
\author{Cl\'{e}ment Surville, Judit Szul\'{a}gyi, Lucio Mayer}
\affil{Institute for Computational Science, University of Zürich, Winterthurerstrasse 190, 8057 Zürich, Switzerland.}
\and
\author{Maria Schönbächler}
\affil{Institut für Geochemie und Petrologie, Departement Erdwissenschaften, Clausiusstrasse 25, 8092, Zürich, Switzerland.}

\begin{abstract}

Chondrules are crystallised droplets of silicates formed by rapid heating to high temperatures ( $> 1800$ K) of solid precursors followed by hours or days of cooling. Dating of chondrules is consistent with the formation timescale of Jupiter in the core-accretion model (1-4 Myrs). Here we investigate if the shocks generated by a massive planet could generate flash heating episodes necessary to form chondrules using high resolution 2D simulations with the multi-fluid code RoSSBi. We use different radiative cooling prescriptions, i.e. different cooling rates and models, and vary planet mass, orbit and disk models. For long disk cooling rates ( $> 1000$ orbits) and a massive protoplanet ( $> 0.75$ $M_{\jupiter}$), we obtain hot enough temperatures for chondrule formation, while using more realistic thermodynamics is not successful in the Minimum Mass Solar Nebula (MMSN) model. However, sufficient flash heating is achieved by a Jupiter mass planet in a 5 times more massive disk, which is a conceivable scenario for the young solar nebula and exoplanetary systems. Interestingly, a gap-forming massive planet triggers vortices, which can trap dust, i.e. chondrule precursors, and generates a high pressure environment that is consistent with cosmochemical evidence from chondrules. A massive gas giant can thus, in principle, both stimulate the concentration of chondrule precursors in the vicinity of the shocking regions, and flash-heat them via the shocks.

\end{abstract}


\section{ Introduction }
\label{Sect_Intro}

	Chondrules are one of the main components (up to ${80 \%}$, \cite{Jones2012}) of chondrites, which are considered to represent building blocks of terrestrial planets and giant planet cores. For this reason, they are important and bear essential information on the first stages of planet formation. Chondrules are crystallised droplets of mainly silicate melt and can be subdivided into various groups depending on their texture, mineral content, and chemistry (\cite{Jones2012} and references therein). Experiments and chemical modelling \citep{1988mess.book..660H, 1996cpd..conf..187L, Hewins2005} show that each chondrule group requires specific formation conditions. Nevertheless, general characteristics can be derived. Chondrules were heated to high temperatures ($> 1800-2250$ K) for short periods -- in flash heating events. Their textures also imply that they experienced relatively fast cooling from 0.5 to 3000 K.h$^{-1}$. Other constraints can be derived from their chemical composition. For example, a high gas-to-dust ratio is required for chondrule formation in the protoplanetary disk to explain their high Na content, a volatile element that would otherwise have been lost from the system \citep{Alexander2008}. 

	Over the last few decades, several processes have been suggested for the formation of chondrules. This includes chondrule formation in (i) impact splashes during collision of planetary bodies \citep{Dullemond2014, Dullemond2016a, Lichtenberg2018, Sanders2012}, or (ii) lightning bolts in the solar nebula \citep{Horanyi1995}, (iii) nebular bow shocks, generated for example by planetesimals \citep{Gong2019}, (iv) general nebular shocks \citep{Desch2002}. Despite, and most likely because of, this rich history and the considerable amount of data obtained from chondrules in meteorites, the mechanisms of chondrule formation are still widely debated. Shocks have been suggested to offer favourable conditions for chondrule formation in terms of heating \citep{Iida2001, Desch2002, Morris2010}. Such shocks create high enough temperatures to melt chondrule precursors and form chondrules if shock velocities reach about 8 km.s$^{-1}$ \citep{Morris2016}. 

	Complementary to cosmochemical evidence, recent observations by the Atacama Large Millimeter Array (ALMA) radio telescope identified large gaps in protoplanetary disks. One of the main hypotheses to explain these gaps is the presence of forming planets in the disk that clean their orbits of gas and dust \citep{Andrews2016, Liu2018}. A planet moving in the disk causes nebular shocks, in which temperature and pressure are elevated. The 1D shock model of \cite{Stammler2014} indicates that in most cases the planet-generated shocks are not energetic enough to induce sufficient high temperatures. However, these simulations do not include the effects of optical densities or the structure of the disk, which 2D simulations do. The model also uses a simpler description of the 1D shock, and thus cannot describe the whole complexity of shock temperature. Mass, momentum, and energy are transported differently through the shock front, if additional 2D or 3D dimensionality is applied. Moreover, the non-linear components of the dynamical equations affect the shock, and in particular the temperature. Linear analysis is able to predict the physical variables in the waves excited by the planet \citep{Rafikov2002}, but barely the temperature because linear models are usually isothermal. Non-linear waves are typically produced, when the planet mass increases above 10 Earth masses. The resulting shock properties cannot be predicted by analytical models \citep{Richert2015}. Therefore, numerical model in 2D or 3D are necessary to accurately capture the dynamics and thermodynamics of the system.

	To address the origin of chondrules, here for the first time, global, 2D disk hydrodynamical simulations are applied to test whether the presence of Jupiter in the early solar system can trigger shocks in the disk that are able to produce the required conditions to form chondrules. In this article, a parameter space with varying planet mass, semi-major axis, and cooling methods is explored to assess the shocks caused by the presence of Jupiter and their ability to generate high enough temperatures to melt chondrule precursors.

\section{ Numerical technique and physical model }
\label{Sect_Methods}

	The RoSSBi code developed by C. Surville \citep{Surville2015, Surville2016, Surville2019} is used to investigate the effects of a Jupiter-like planet on a disk. It is an accurate second-order finite volume scheme, designed to solve the 2D compressible fluid equations on a polar grid, including dusty flows. The ability of the RoSSBi code to capture shocks with very low diffusion is perfectly adequate for this study, because it was developed to preserve shocks and compressibility effects.

	The presence of a Jupiter analogue in the disk is included to simulate the influence of such a planet on the disk. As typical in planet-disk interaction simulations, a gravitational softening length of $0.6$ times the local disk scale height is used to avoid the singularity of the potential at the planet location, and to mimic the effect of the disk's vertical structure on the gravitational potential. The motion of the planet is then integrated with a LeapFrog method, and the gravitational influence of the disk on the planet is computed in some of the runs. Runs with a fixed eccentricity were also conducted to assess the effects of this parameter. Consequently, momentum and energy can be exchanged between the disk and the planet, leading to planet migration and eccentricity variations, for instance.

	The conditions in the solar protoplanetary disk are investigated using the Minimum Mass Solar Nebula (MMSN) model \citep{Hayashi1981}. We assume the unperturbed gas surface density $\sigma_0(r) = 1700 \times (r/1 \: au )^{-3/2}$ g.cm$^{-2}$ with $r$ the distance from the Sun. This typical axisymmetric profile is referred to as the background profile. The background temperature $T_0(r)$ varies in the disk as $r^{-1/2}$, and the pressure is given by the ideal gas law prescription, $P_0(r) \propto \sigma_0(r) T_0(r)$. We define a reference radius, $r_0$, given in astronomical units, which sets the location of the disk domain we simulate. The absolute pressure value (and thus the temperature) is given by setting the vertical aspect ratio of the disk at $r_0$. This disk scale height at $r_0$ is given by $\sqrt{P_0/\sigma_0}/\Omega_k = 0.05 r_0$, where $\Omega_k(r)$ is the Keplerian angular velocity imposed by the Sun.

	Finally, the thermodynamics of the gas is solved via the energy equation. By default, the code RoSSBi adopts the adiabatic assumption, which we used in early simulations. For comparison, we also added and tested two different radiative cooling terms (i) a thermal relaxation (or beta cooling) source term of internal energy and (ii) "kappa" cooling. For thermal relaxation (or beta cooling, Equation \ref{Equ_beta_cooling}) one describes cooling as a relaxation process such that the temperature of the gas tends towards a prescribed background equilibrium temperature $T_0(r)$ at a given rate,

\begin{equation}
\label{Equ_beta_cooling}
 \frac{{\partial T}}{{\partial t}} = - \frac{\Omega_k(r)}{\beta_c} [T - T_0(r)].
\end{equation}

The cooling rate is thus scaled to the local orbital frequency, and the cooling parameter $\beta_c$ is dimensionless. A large value tends towards the adiabatic solution, whereas a small $\beta_c$ value towards the isothermal limit. We focus on the disk midplane, where the gas is optically thick, and cannot radiate efficiently. Therefore, we only use values of $\beta_c >> 1$.
For "kappa" cooling, simulations were run solving an approximate equation that models radiative diffusion plus its transition to the free-streaming limit at low optical depths:

\begin{equation}
\label{Equ_kappa_cooling}
 \frac{{\partial e_i}}{{\partial t}} = - \frac{4 \sigma_B [T^4 - T_0^4(r)]}{\sigma (\tau_R+1/\tau_P) } ,
\end{equation}

where $e_i$ is the internal energy per mass unit, $\sigma$ is the gas surface density, $\sigma_B$ is the Stefan-Boltzmann constant, $T$ is temperature, $\tau_R$ is Rosseland's optical depth, and $\tau_P$ is Planck's optical depth. This model of thermodynamics is proposed by \citep{Stamatellos2007} in smoothed particle hydrodynamics (SPH) simulations of self-gravitating disks. This simplified model based on local conditions compares well to full radiative models. For the first time, this method is implemented here in a finite volume method.

	The vertical optical depth $\tau$ (Rosseland or Planck) is obtained from the mean opacity, either Rosseland or Planck, and is related to the opacity $\kappa$ by : 

\begin{equation}
\label{Equ_optical_depth}
 \tau = \int \kappa \rho dz,
\end{equation}

with $\rho$ the gas volume density, and the integral running to infinity.

	Because we work with a 2D disk model, we impose a thermal equilibrium in altitude, i.e. temperature is constant with z. As a result, the opacities are assumed constant vertically. Using the Rosseland mean opacity in the previous equation (Equation \ref{Equ_optical_depth}) gives the optical depth $\tau_R$ , similarly using the Planck opacity gives the Planck optical depth $\tau_P$ . Using these two optical depths reflects the smooth transition between purely free-streaming photons (black body radiation) and fully diffusive radiation that the model constructs. However, it is assumed that $\kappa_P = \kappa_R = \kappa$, in the model. The value of the mean opacity kappa follows the parametrization as function of the temperature given by \cite{Bell1994}, and we apply the coefficients given in \cite{Mercer2018}. In particular, we neglect the sharp transitions in opacity at $T > 2000$ K for simplicity because it occurs at temperatures higher than needed for melting of chondrule precursors.
Finally, the internal energy ($e_i$) per unit mass depends on temperature by the following relations :

\begin{equation}
\label{Equ_internal_energy}
 e_i = \frac{cs^2}{\gamma - 1},
\end{equation}

with $\gamma$ the adiabatic index of the gas (here, $\gamma = 1.4$). The sound speed squared, $cs^2$, is obtained from :

\begin{equation}
\label{Equ_sound_speed}
 cs=\sqrt{ \frac{k_B T}{\mu m_H} },
\end{equation}

with $k_B$, Botlzmann constant, $\mu$ the mean molecular weight of the gas (here $\mu = 2.34$) and $m_H$ the mass of an atom of hydrogen.

	This approximate method has two main advantages. It covers a range of cooling conditions from radiative diffusion, when $\tau >>1$, to the black body radiation, when $\tau << 1$, with a smooth and continuous transition between these extreme regimes. The second advantage is to avoid the dependence of the flux of radiation on the local gradients, which reduces computing complexity. The cooling conditions are applied locally as in the thermal relaxation technique. However, the cooling rate for kappa cooling depends on both the local temperature and the local optical depth.
To draw a comparison with the beta-cooling model, we can write the radiative diffusion equation (Equation \ref{Equ_kappa_cooling}) in terms of temperature in the form:

\begin{equation}
\label{Equ_kappa_cooling_1}
 \frac{{\partial T}}{{\partial t}} = - \frac{\Omega_k(r)}{\beta_{\kappa}} [T - T_0(r)].
\end{equation}

Here $\beta_{\kappa}$ is an effective cooling rate depending on temperature, surface density and optical depth, and is expressed in number of orbits in order to allow direct comparison with $\beta_c$. By substitution in Equation \ref{Equ_kappa_cooling}, we obtain the equivalent cooling rate $\beta_{\kappa}$ of this model :

\begin{equation}
\label{Equ_kappa_cooling_2}
\begin{split}
 \beta_{\kappa} = & \frac{\sigma (\tau_R+1/\tau_P) }{16 \sigma_B T_0^3(r)} \frac{k_B \Omega_k(r)}{\mu m_H (\gamma-1)} \\
 & \times \left[1+ \frac{3}{2}\frac{\Delta T}{T_0(r)} + \left[\frac{\Delta T}{T_0(r)}\right]^2 + \frac{1}{4}\left[\frac{\Delta T}{T_0(r)}\right]^3 \right]^{-1} ,
\end{split}
\end{equation}

where $\Delta T = T - T_0(r)$.

\begin{figure*}
	\begin{center}
	\includegraphics[width=15cm, trim=0mm 0mm 0mm 0mm, clip=true]{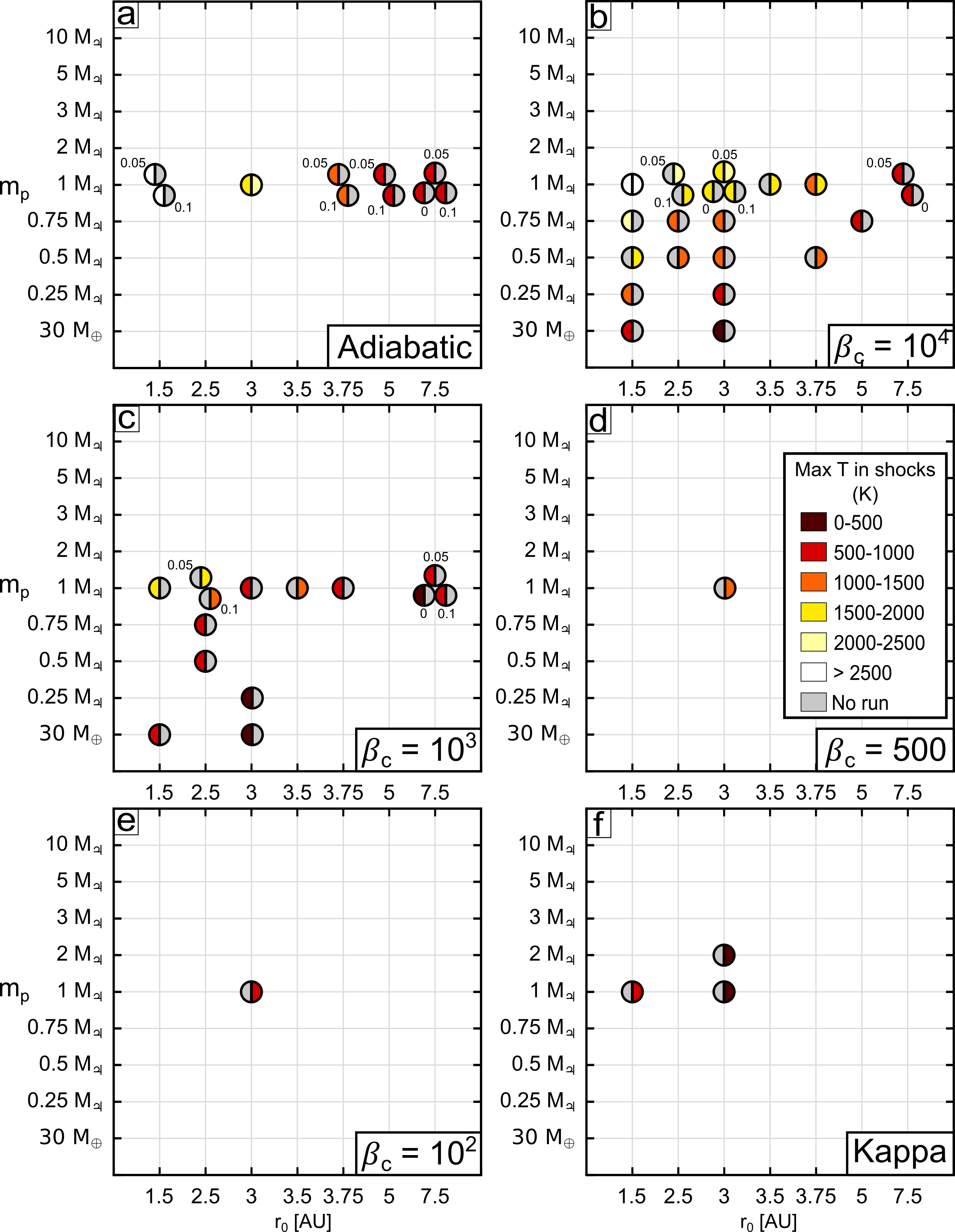}
	
	\caption{\label{Fig_1} Summary plots of explored parameter space. Each plot has $r_0$ in au on the x axis and mp in units of Jupiter mass ($M_{\jupiter}$) and Earth mass ($M_{\bigoplus}$) on the y axis. Each dot is separated in two parts, left: fixed eccentricity, and right: moving planet. Greyed out parts refer to the cases where no simulations were ran for the associated option. Grouped dots represent variations of initial eccentricity, with the associated eccentricity indicated next to each dot. Where no indication is given, the initial eccentricity value is $0.05$. The colour scale represents the maximum temperature range achieved in the simulation with the given parameters. Each panel represents different cooling modes and parameters with a. adiabatic case; b., c. ,d., e. thermal relaxation with $\beta_c = 10^4$, $10^3$, $500$, and $100$, respectively; f. cooling by approximation of radiation diffusion ("kappa" cooling). }
   \end{center}
\end{figure*}

	Several parameters were investigated in this study. They are summarized in Figure \ref{Fig_1}. Two of the main parameters are $r_0$, which is the semi-major axis, the distance between the planet and the Sun, given in au and the mass of the planet. In the code, the mass is given in units of solar mass ($M_{\astrosun}$). Here we also use Jupiter mass ($M_{\jupiter}$) and Earth mass ($M_{\bigoplus}$) as illustrative units to describe planet masses. The effects of the orbit eccentricity and cooling methods are also assessed (see Figure \ref{Fig_1}). In the resulting maps, gas density is represented by the symbol $\sigma$. Pressure and density are normalised to the background values and are thus presented without units. Snapshots of the simulation are acquired and dumped every 2 disk rotations. 

\begin{figure*}
	\begin{center}
	\includegraphics[width=18cm, trim=0mm 0mm 0mm 0mm, clip=true]{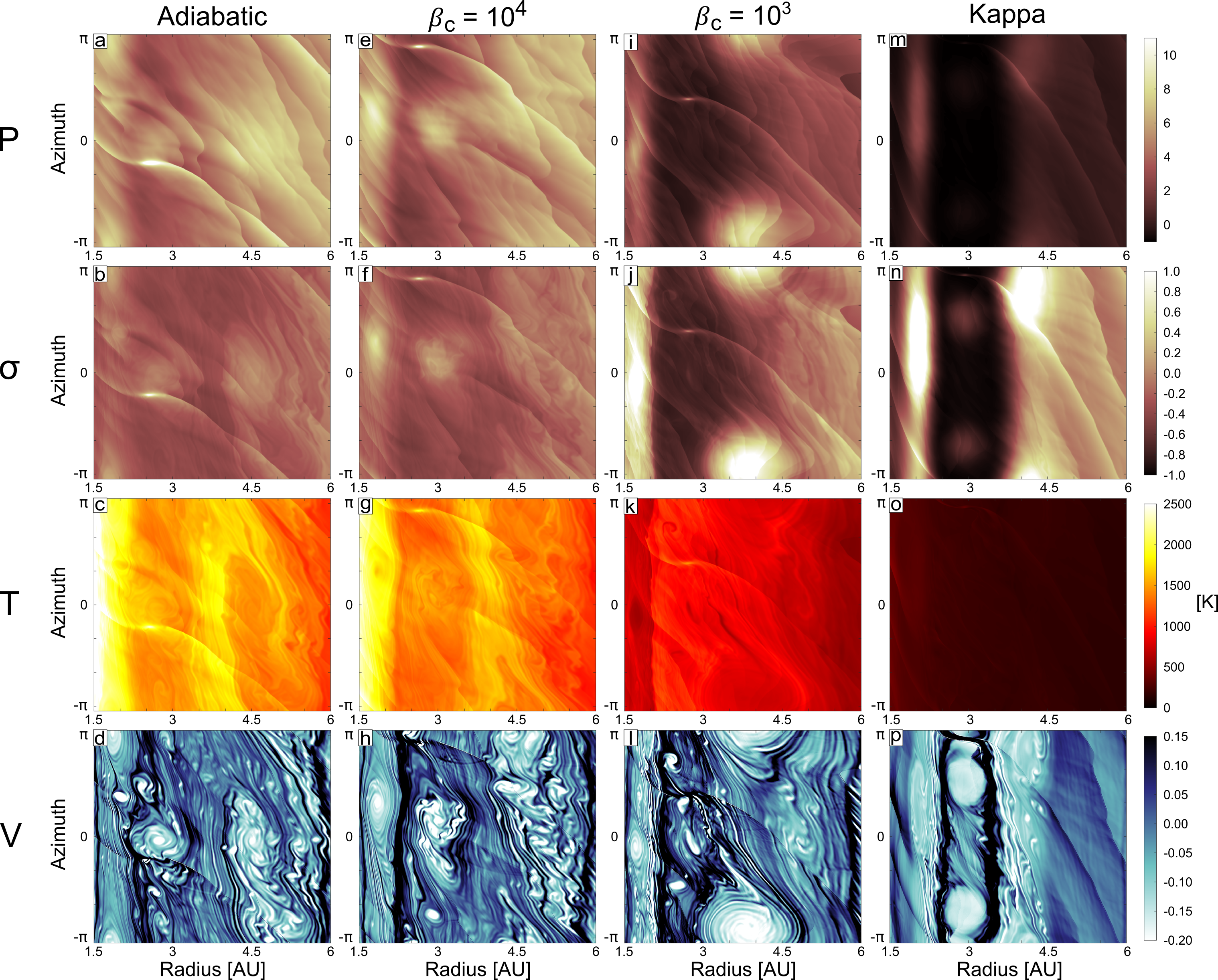}
	
	\caption{\label{Fig_2} Results of simulations for different cooling rates, $\beta_c$. The first row of the panels represents gas pressure (P), the second gas density ($\sigma$), the third temperature (T, in K) maps, and panels on the fourth row show vorticity (V). Pressure and density are normalised to the background values, and are thus dimensionless. V is represented by the dimensionless Rossby number. Results obtained after 300 disk rotations, for a Jupiter mass planet orbiting at 3 au with an eccentricity of 0.05}
   \end{center}
\end{figure*}

\section{ Numerical simulations setup}
\label{Sect_Setup}

	Simulations presented here were run without dust because its addition greatly increases computing time. However, dust grains with the size range of putative chondrule precursors (nm - $\mu$m size), are strongly coupled to the gas \citep{Surville2019}. The simulations start with a homogeneous distribution of gas. A temperature profile is defined across the disk using a power law that corresponds to typical models of the minimum mass solar nebula. In some simulations, the planet starts with a specific eccentricity and is allowed to migrate under the influence of the disk afterwards (moving planet, Figure \ref{Fig_1}). Simulations with a fixed planet orbit were also performed to assess the effects of eccentricity on shocks and disk structure. Three values were investigated, $e = 0$, $e = 0.1$, and $e = 0.05$.

	The resolution of the inviscid simulations presented in this work is $N_r = 1024$ and $N_{\theta} = 1024$ cells, in radial and azimuthal directions. Boundary conditions were handled using ghost cells according to the methods described by \cite{Surville2015}. Different cases of disk cooling were investigated. Some simulations were run with adiabatic conditions. Several thermal relaxation runs were investigated with $\beta_c$ values ranging between $10^2$ and $10^4$ orbits. Finally, we performed runs using the second more realistic cooling model, i.e. with the "kappa" cooling equation.

	The effect of planetary mass was assessed by selecting a range of values from 30 $M_{\bigoplus}$ (approximate mass of Jupiter's core before it accreted its gas, see e.g. \cite{Safronov1991}, and \cite{Guillot1997}), to 1 $M_{\jupiter}$, the mass of a fully formed Jupiter.

\section{ Results }
\label{Sect_Results}

\subsection{ Overview }
\label{Sect_Results_overview}

	A large suite of simulations was run, 74 in total, varying cooling models, planet semi-major axis, orbital eccentricity, and in a few cases, disk mass. The main results of these simulations, such as gas pressure, density, flow velocity, and vorticity are presented as maps in polar coordinates. They are used to characterise flow properties in particular in the region near the planet, where potential chondrule-forming shocks are triggered. In all maps, the x-axis represents the radial distance from the Sun, and ranges from 0.5 to 2.5 $r_0$. The y-axis represents the azimuth in the disk ranges from 0 to 2$\pi$ radians.

\begin{figure*}
	\begin{center}
	\includegraphics[width=18cm, trim=0mm 0mm 0mm 0mm, clip=true]{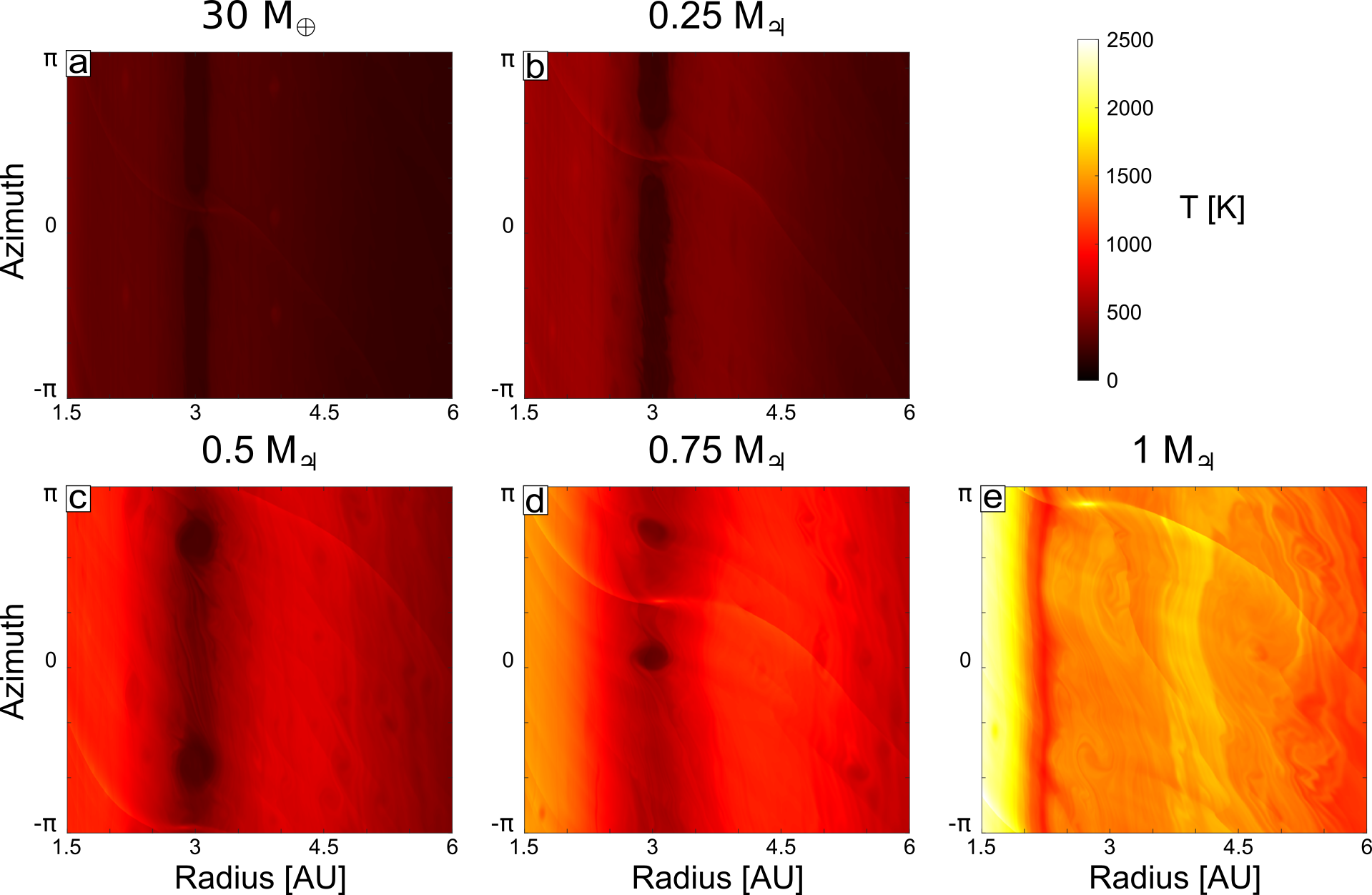}
	
	\caption{\label{Fig_3} Maps of temperatures acquired for runs with varying planetary mass. Fixed parameters are $r_0 = 3$ au; $e = 0.05$; $\beta_c = 10^4$; snapshot at 354 orbits. }
   \end{center}
\end{figure*}

\begin{figure*}
	\begin{center}
	\includegraphics[width=18cm, trim=0mm 0mm 0mm 0mm, clip=true]{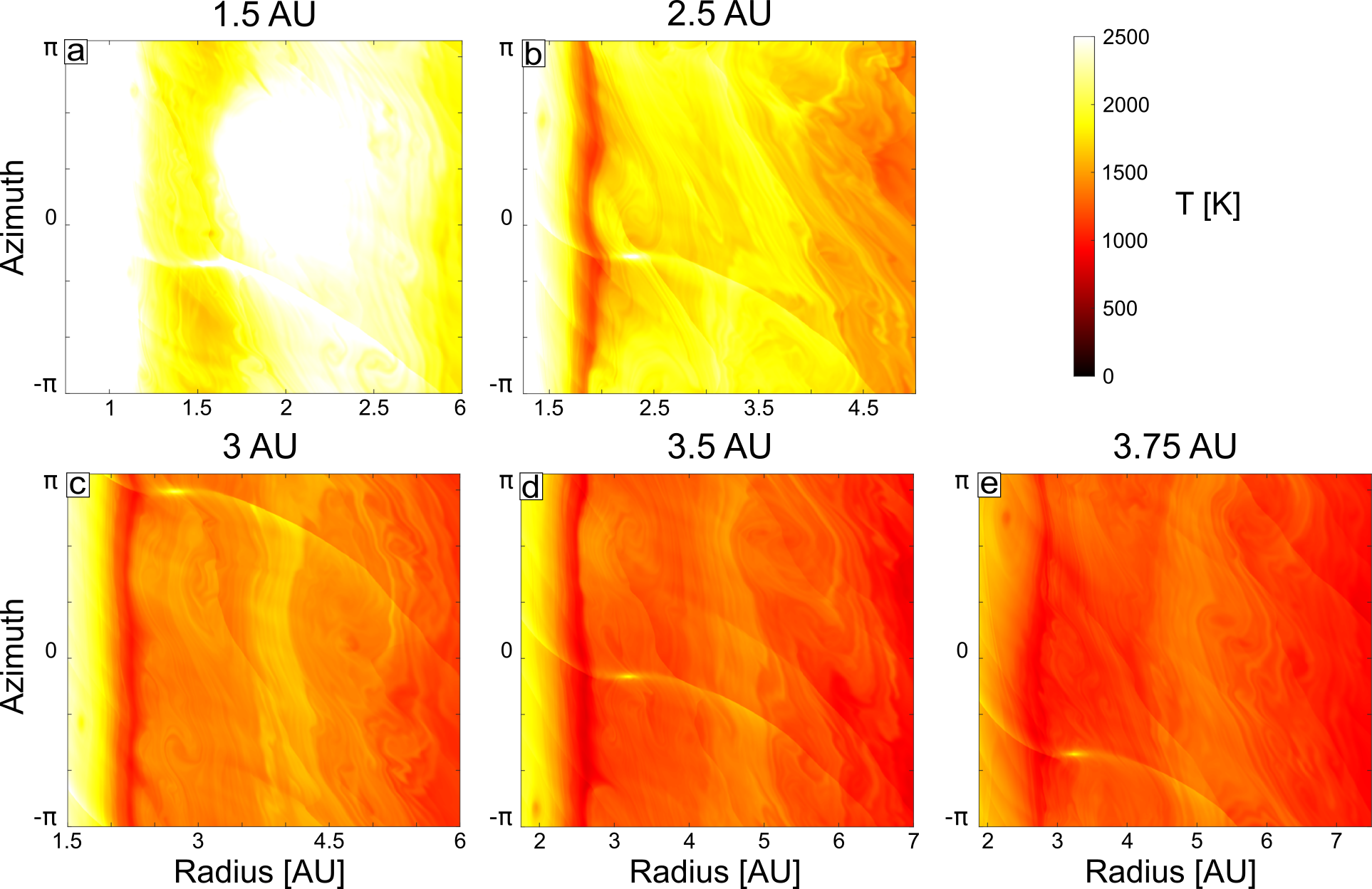}
	
	\caption{\label{Fig_4} Maps of temperatures acquired for runs with varying semi-major axis values, for a Jupiter mass planet. Fixed parameters are $e = 0.05$; $\beta_c = 10^4$; snapshot at 354 orbits }
   \end{center}
\end{figure*}

	Figure \ref{Fig_2} shows the effect of the disk cooling on the disk evolution with a Jupiter-mass planet orbiting at 3 au. The effect of the mass of the planet is shown Figure \ref{Fig_3}. Finally, we show \ref{Fig_4} the influence of the orbital distance of the planet on the disk temperature.

	We now describe the general behaviour observed in the simulations. One of the main features present in all runs is the decrease of gas and dust density and low pressure on each side of the planet orbit as the planet travels through the disk. This reflects the gap that is carved by the planet and centred on its orbit. The width of the gap region depends on the planetary mass, ranging from 0.3 au for a 30 $M_{\bigoplus}$ planet (Figure \ref{Fig_3}, panel a.) to up to 1 au for a 1 $M_{\jupiter}$ planet (Figure \ref{Fig_3}, panel e.), when it is fully formed. It also depends on the local disk scale height. The magnitude of the depletion of gas density and pressure also increases with planetary mass. Consequently, dust density also should also decrease accordingly.

\begin{figure}[t]
	\begin{center}
	\includegraphics[width=8.5cm, trim=0mm 0mm 0mm 0mm, clip=true]{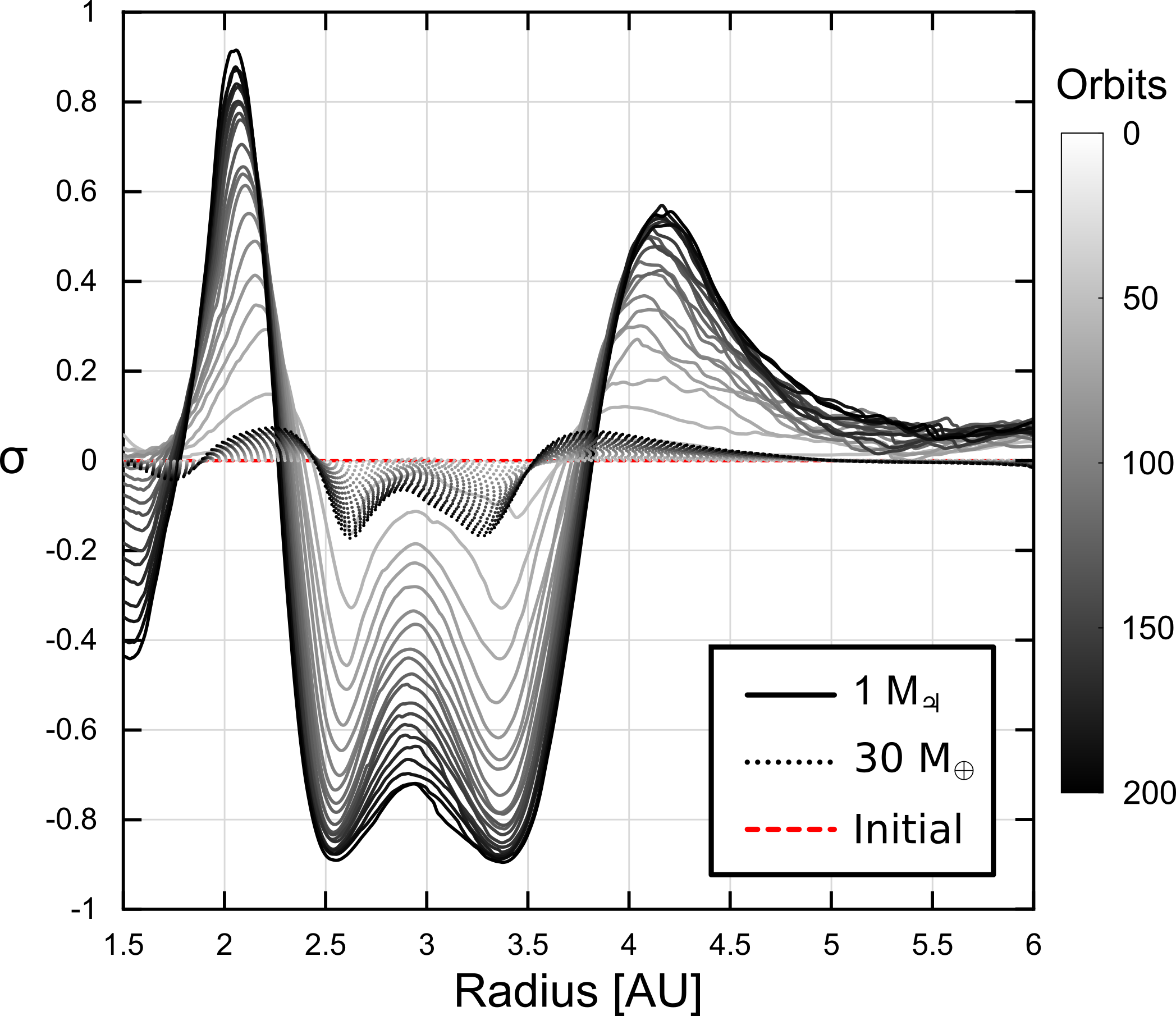}
	
	\caption{\label{Fig_5} Temporal evolution of azimuthal average of surface density (density normalised to background density) for disks with a Jupiter mass planet (continuous lines) and a 30 Earth mass planet (dashed lines) from 0 to 200 orbits. The number of orbits is given by the line colour, going from light grey (0 orbit) to black (200 orbits), see scale. Fixed parameters are $r_0 = 3$ au; $e = 0.05$; "kappa" cooling, with a moving planet. The gap region is centred around 3 au and characterised by low gas density. Higher densities are found on each side of the gap and correspond to the region where vortices form. }
   \end{center}
\end{figure}

	Gaps are less well defined in the adiabatic and $\beta_c = 10^4$ (Figure \ref{Fig_2}, panels a.-h.) cases compared to those with stronger cooling (Figure \ref{Fig_2}, panels i.-p.). In the former cases, the energy deposited by the shock takes the form of thermal energy instead of kinetic energy. It causes higher temperatures in the disk, but also less well-defined gap structures. It is well known that isothermal simulations create deeper gaps \citep{Les2015}, which is coherent with these results. The speed at which the gap forms depends on the mass of the planet, but is generally relatively fast (Figure \ref{Fig_5}). For a 30 $M_{\bigoplus}$ planet, the gap forms progressively and is not well marked before around $t = 60$ rotations, while for planets of 1 $M_{\jupiter}$, a significantly sharper and deeper gap can be already be carved in $t < 20$ rotations.
Density and pressure enhancements are observed in the disk both on the inner and outer side of the gap (Figure \ref{Fig_2}). The highest pressure and gas densities occur close to the planet and in the shocks with pressures reaching more than 10 times that of the background gas for a planet of 1 $M_{\jupiter}$. Pressure is also enhanced in the vortices (e.g. Figure \ref{Fig_2}, panels j. and l.).

\begin{figure}[t]
	\begin{center}
	\includegraphics[width=8.5cm, trim=0mm 0mm 0mm 0mm, clip=true]{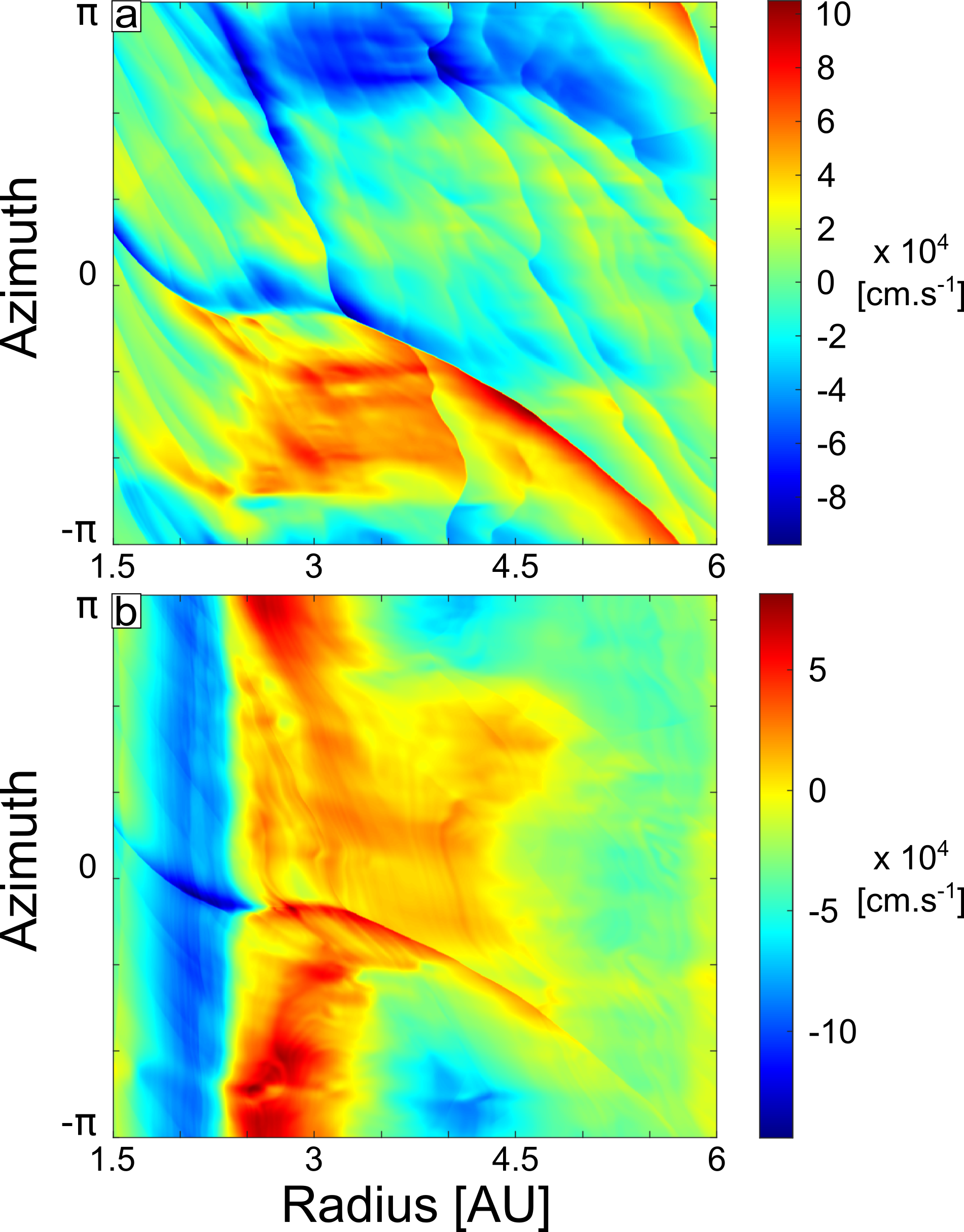}
	
	\caption{\label{Fig_6} Maps of radial, and azimuthal components of the velocity field (top and bottom, respectively) relative to the background Keplerian velocity. Snapshot at 300 orbits in a simulation with a Jupiter mass planet, orbiting at 3 au, with an eccentricity of 0.05, and a disk cooling rate $\beta_c = 10^4$. }
   \end{center}
\end{figure}

	The velocity of the gas was tracked across the disk. We show Figure \ref{Fig_6} the radial and azimuthal components of the velocity field (top and bottom, respectively), whose magnitude is given in cm.s$^{-1}$. They are calculated relative to the local Keplerian velocity, hence, the presence of negative values in the Figures. 
Azimuthal velocities (Figure \ref{Fig_6}, bottom) show a similar distribution in most runs, with an area of high positive velocities just outside of the planet's orbit and a region of high negative velocities inside the planetary orbit. This structure, associated to the planet gap, develops early, within twenty orbits, and is usually maintained until the end of the run. Velocities vary between -1.2 and 1.2 km.s$^{-1}$ with the higher spread found in simulations with higher planetary masses, and colder overall temperatures. Radial velocities (Figure \ref{Fig_6}, top) are more complex and the structures change widely between runs and even between two snapshots in the same run. Common features identified are high positive velocities in the shock accompanied by negative velocities in the gas in front of it. They can vary between -1.5 and 1.5 km.s$^{-1}$. The planetary mass has a major effect on the radial velocities, resulting in disks with more disturbances for higher planet mass. However, in low mass cases, the changes in velocity are essentially confined to the shock and its direct surroundings. This is in agreement with the presence of non-linear inertial waves excited by massive planets \citep{Richert2015}.

	While we predicted azimuthal velocities to be dominant, and this certainly holds true close to the planet, our simulations demonstrate that radial velocities cannot be ignored. In particular, when far from the planet, the radial velocity component of the shock becomes more important than the azimuthal velocity component, and can reach values above 1.5 km.s$^{-1}$. Hence, it plays an important role in the generation of the heat available to melt chondrule precursors. We note that even the highest flow velocities achieved in the simulated shocks (maximum 1.7 km.s$^{-1}$) are not sufficient per se to generate enough heat to melt chondrules \citep{Morris2010}. However, these are shock propagation velocities relative to the background flow, which is already heated due to the work of pressure forces. This explain why in many of our runs the temperature can still reach 1800 K and above, which satisfies the melting conditions for chondrules.

	Therefore, to allow for comparison of our results to constraints obtained from chondrules, we directly use the temperature T, derived as in Equation \ref{Equ_sound_speed}. Note that $r_0$ is one of the main parameters that controls temperature, because of the radial variation of the background temperature in the disk. The smaller $r_0$ is, the higher the temperature as shown in Figure \ref{Fig_1} and Figure \ref{Fig_4}.

	Temperatures vary widely in space and time, even in single simulations. As the shock caused by the planet propagates through the disk, the compression of the gas causes the temperature to increase, resulting in higher temperatures in the planet generated spiral wakes. In the adiabatic cases, temperature builds up throughout the simulation, resulting in the hottest disks for a given $r_0$ compared to other cooling regimes. Applying cooling causes the temperatures to rise slower and to reach lower temperatures. In this case, the system can finally reach a thermal equilibrium. Within individual runs and time steps, temperatures also vary. In most cases, temperatures are higher close to the planet, in the shock, and on the inner edge of the gap. Areas of higher temperatures also develop beyond the gap, in regions of higher density, pressure, and vorticity. In the cases with the most intense cooling ($\beta_c = 10^2$, "kappa" cooling), temperatures in the outer part of the disk are distributed differently as seen in Figure \ref{Fig_2}, panel k., with a cold inner disk, higher temperatures in the region of low density and gas pressure and low temperatures in vortices and in the outer disk.

	Finally, we produced maps of the vorticity of the gas, given as the Rossby number $Ro=||\vec{\nabla} \times [\vec{V}-\vec{V_0}(r)]||/[2 \Omega_k(r)]$, with $\vec{V_0}(r)$ the disk background velocity. As with density and pressure, low vorticities are observed in a band of 1 au width (Figure \ref{Fig_2}) centred on the planet's orbit. It is notable, however, that two vortices form ahead and behind the planet in this band with cases with realistic "kappa" cooling (Figure \ref{Fig_2}, panel p.). In most cases, the formation of a large scale vortex occurs at the outer edge of the gap. It is a result of a Rossby Wave instability (RWI) developing at the gap edge. This development of the RWI instability at the gap location has been observed in many models evolving massive planets in disks, i.e. for planets with mass larger than 30 Earth masses \citep{DeVal-Borro2007, Fu2014, Zhu2014}. The RWI saturates into vortices very quickly, only 20 disk rotations after the gap opening. In the runs with a Jupiter mass planet, the formation of a vortex also occurs at the inner edge of the gap. 

\subsection{ Successful and unsuccessful flash heating: the critical role of cooling }
\label{Sect_Results_cooling}

	Three cases for cooling have been investigated in this study. In the first case, the disk is adiabatic and no thermal relaxation occurs. For the first set of runs with cooling, thermal relaxation has been added to the simulation. Values of $\beta_c = 10^2$, $\beta_c = 5 \times 10^2$, $\beta_c = 10^3$ (Figure \ref{Fig_2}, panels i.-l.), and $\beta_c = 10^4$ (Figure \ref{Fig_2}, panels e.-h.) disk rotations were investigated. Finally, a more realistic method, using the "kappa cooling" model, was implemented and its effects on simulations assessed (Figure \ref{Fig_2}, panels m.-p.).

	In adiabatic simulations (Figure \ref{Fig_2}, panels a.-d.), the shock caused by Jupiter keeps heating the disk and the temperature increases continuously without reaching thermal equilibrium. This situation corresponds to highly inefficient cooling because the high gas opacity yields a radiative diffusion time that is much longer than the local orbital time (as it is also the case in the relaxation runs with highest $\beta_c$). While temperatures to melt chondrule precursors are easily achieved in these cases, the high temperatures over large areas of the disk are an issue for the subsequent steps of chondrule formation i.e. the relatively short cooling time scales of chondrules. It is particularly problematic in simulations with a planet close to the Sun where the temperatures can reach several thousands of K in the disk (Figure \ref{Fig_4}). This exceeds the maximum melting temperature for chondrules by far, likely even vaporizes the melt, and does not allow chondrule melts to cool down sufficiently fast to re-crystallise. This scenario is thus deemed unlikely. For runs with $\beta_c = 10^4$ (Figure \ref{Fig_2}, panels e.-h.), disk thermal structures are similar to those observed in the adiabatic case, with the development of high temperatures in the inner disk, and outside of the planet's orbit in shocks. 

	In simulations with $\beta_c = 10^3$ orbits (Figure \ref{Fig_2}, panels i.-l.), however, the thermal structure of the disk deviates strongly from those described above. While in the early stages of the simulations, temperature increases similarly, albeit slower, to the $\beta_c = 10^4$ case, the structure changes after around 100 rotations. Temperatures in both the vortex and the inner disk are much lower than in other simulations with otherwise identical parameters, although in specific cases, temperatures in shocks can still be high enough to reach the peak temperatures needed to form chondrules (Figure \ref{Fig_1}). However, in cases where $\beta_c = 10^3$, the high temperatures are closer to the planet and do not reach the vortices, rendering this scenario unlikely for chondrule formation. Similar structures are observed in the cases with $\beta_c = 10^2$, and $\beta_c = 5 \times 10^2$, but with temperatures decreasing with $\beta_c$ (Figure \ref{Fig_1}). The most likely scenario to form chondrules in a disk with thermal relaxation is thus $\beta_c = 10^4$, excluding the adiabatic case. Whether these cooling timescales are realistic, was tested by the last set of runs with radiative diffusion ("kappa cooling"). 

\begin{figure}[t]
	\begin{center}
	\includegraphics[width=8.5cm, trim=0mm 0mm 0mm 0mm, clip=true]{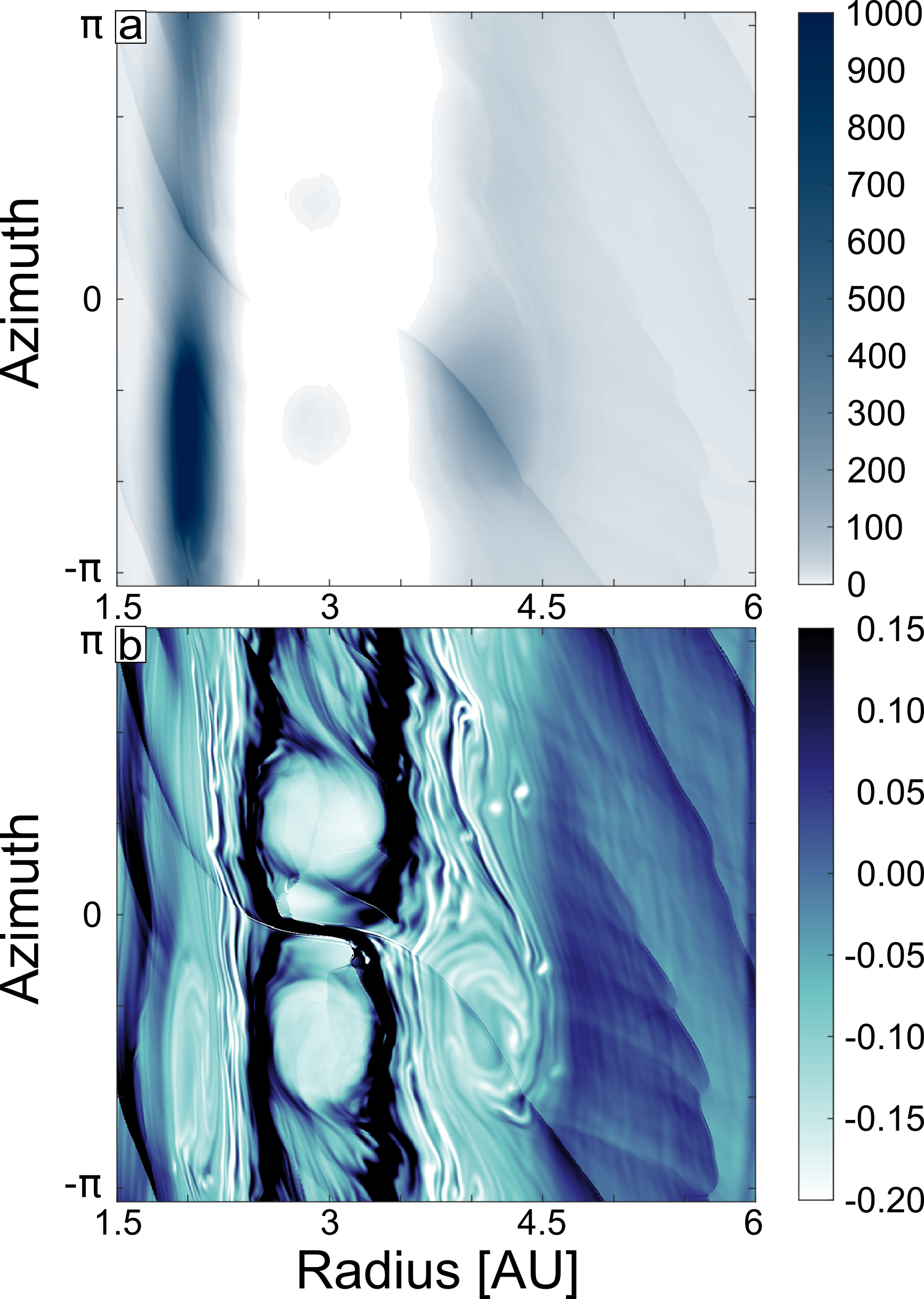}
	
	\caption{\label{Fig_7} Top: map of the equivalent cooling rate, $\beta_{\kappa}$, obtained when using the "kappa" cooling model. Bottom: map of the vorticity of the gas, given as the local Rossby number. The vorticity map is given for reference to show the location of shocks and vortices. Snapshots at 365 orbits in a simulation with a Jupiter mass planet, orbiting at 3 au, with planet migration. The "kappa" cooling model is used for the gas thermodynamics. }
   \end{center}
\end{figure}

	Simulations ran using the "kappa cooling" approximation to radiation diffusion (Figure \ref{Fig_2}, panels m.-p.) and with the conditions according to the model of the MMSN do not achieve the temperatures required to melt chondrule precursors, even for massive planets as close as 1.5 au. These runs display different cooling timescales across the disk (Figure \ref{Fig_7}, top). This is due to the varying optical depth through the disk. Typically, while cooling is very fast in the gap (0 to 1 orbit), it is slower in vortices. In the outer vortex, $\beta_{\kappa}$ can reach up to 300 orbits and over 1000 orbits in the vortex observed in the inner part of the disk. $\beta_{\kappa}$ is also higher along the shock front (Figure \ref{Fig_7}) because the high compression there increases the local gas density and thus optical depth. This results in $\beta_{\kappa}$ increasing by up to 100 inside and outside of vortices relative to regions not affected by the shock front. Yet, these values of $\beta_{\kappa}$ are still lower than the $\beta_c = 10^3$ and $\beta_c= 10^4$ used in most simulations ran with thermal relaxation, which explains the lower temperatures achieved. 

	The MMSN nebular model contains only the minimum mass necessary to form the solar system planets. It has often been suggested that the real mass of the solar nebula might have been higher \citep{Lissauer1987, Lissaue1993}. Therefore, simulations were also ran using 5 times the MMSN for the initial mass of the solar protoplanetary disk. In these simulations, temperatures can increase sufficiently to melt chondrule precursor both outside and inside the orbit of a massive planet with the planet orbiting close to the Sun ( $< 2$ au). While this is promising, the more massive nebula causes a different set of problems as discussed in Section 5.

\subsection{ Dependence of shock heating on planetary mass and semi-major axis }
\label{Sect_Results_orbit}

	Planetary masses ranging from the alleged mass of Jupiter's core (30 $M_{\bigoplus}$) up to 1 $M_{\jupiter}$ were investigated. Figure \ref{Fig_3} shows the result of simulations with a fixed set of parameters with only mp varying. It illustrates that higher T can be reached with more massive planets. Figure \ref{Fig_4} shows 5 runs with $r_0$ being the only changing factor. The results show that the proximity of the planet to the Sun plays an important role. Even for a Jupiter-mass planet, high shock temperatures can only be reached within 3 au of the Sun in most cases, except when adiabatic or close to adiabatic (Figure \ref{Fig_1}). While sufficiently high peak temperatures ($> 1880$ K) for chondrule formation are achieved in some simulations with $r_0 = 1.5$ au, the disk is also very hot so close to the Sun, with gas temperatures outside of the shocks reaching temperatures well over 2000 K. In all cases for this location the background temperature in the gas and dust is too high to allow chondrules to cool and crystallise at the rates required to achieve their textures. Chondrule formation temperatures are only reached in simulations with planets of masses $> 0.5$ $M_{\jupiter}$. For a planet of 0.5 $M_{\jupiter}$, the semi-major axis was only 1.5 au, thus very close to the sun.

	In summary, the most likely location for chondrule formation is between 3 and 2.5 au according to our simulations. However, even at these distances, high enough temperatures in shocks can only be reached when the planet has nearly reached its final mass (75 \% of the mass of Jupiter). Please note that even when the planet is massive enough to generate the required peak temperatures, they are only achieved after a significant number of orbits. At 1.5 au, temperatures caused by the shock can reach over 1800 K within around 50 disk rotations. However, it takes up to 250 orbits to reach these same temperatures in shocks occurring at 3 au.

\section{ Discussion }
\label{Sect_Discussion}

	As evident from previous sections, the parameter space that allows for sufficient flash heating for chondrule formation becomes very small when using our most realistic cooling prescription, the "kappa" cooling model. Moreover, even for simulations with inefficient radiative cooling (adiabatic or $\beta_c= 10^4$ cases), chondrule formation is problematic. The second stage of chondrule formation involves relatively rapid cooling. This is difficult to achieve in these simulations because they show strong heating all over the disk associated with inefficient radiative cooling.

	We also carried out simulation with 5 and 10 times the MMSN because the mass of the solar protoplanetary disk is not very well constrained. The MMSN is a rough extrapolation of the current distribution of mass in the solar system, and is used to convert from the current mass of solids in the solar system to the original mass of the gas assuming solar metallicity. Observations suggest that disk masses can range from 0.5\% to 10\% the mass of the host star, the youngest disks being the most massive \citep{Mohanty2013}. The MMSN model embeds a mass of 1\% the Sun mass within 100 au. It is reasonable to argue that during the first million years, the disk could be a factor of a few more massive.

	The simulations with 5 times the MMSN produce sufficient shock heating for chondrule formation when using the radiative diffusion cooling module. These are the only simulations with this cooling mode, where sufficiently high temperatures for chondrule formation are reached. However, they show other challenges. The interactions between the planet and the more massive disk causes the planet to migrate towards the star within 50 orbits (for 5 MMSN). This is expected given the higher mass of the disk and is also described in systematic studies of giant planet migration with varying disk masses \citep{Malik2015}. When the planet moves closer to the star, around 1.5 au, temperatures in the disk around the planet are much higher than at 3 au. The shocks caused by the planet outside of its orbit can reach temperatures in the range required to melt chondrule precursors. However, in absence of a treatment of dust dynamics and thermodynamics, our simulations show that the temperature within the orbit of the planet remains above 2500 K. This would vaporise the dust in the disk inside the planet and thus prevent chondrule formation. Therefore, the model cannot explain chondrules occurring in ordinary and enstatite chondrites that presumably formed in the inner part of the disk i.e. inside Jupiter (but see below the remaining caveats on the cooling model). Very similar results were obtained in a run with a tenfold increase in gas density (i.e. the fraction of heavy elements) to simulate higher metallicity. 

	Please note that even our most realistic cooling module is still highly simplified. Among the limitations, it cannot take into account the effect of local changes in opacity, for example resulting from dust migration or vaporization following shock propagation. The cooling rate is thus quite uncertain because of the complex flow that we observe in the simulations, with gaps, rims and vortices and shocks, which raise temperatures above the vaporisation temperatures of ices and silicates ($T > 1300$ K). The cooling rate likely increases rapidly after a shock, if the opacity decreases due to vaporisation or dust migration. Hence, assessing the cooling regime for chondrules after shock heating requires improved calculations, which track opacity changes. Such treatment is beyond the scope of the current paper. In addition, it may require a ray-tracing scheme to compute radiation transport more accurately through the highly inhomogeneous flow.

	Despite these difficulties, our results suggest that Jupiter could have played an important role during chondrule formation, even if the planet was not the main cause of it in the Solar System. Jupiter generates a gap, which is ubiquitous in the simulations, independent of the varying thermodynamics of the many different simulations performed for this study. Such a gap is consistent with current early solar system models and astronomical observations such as the ones acquired with ALMA \citep{Partnership2015, Andrews2016}. Meteorite analyses of elements displaying nucleosynthetic anomalies (e.g. Ti and Cr) revealed a dichotomy between two populations of bodies early in the history of the solar system \citep{Trinquier2007, Leya2008}, one in the inner solar system, comprising e.g. of ordinary and enstatite chondrites and one further out characterised by the carbonaceous chondrites and associated iron meteorites. This dichotomy thus arose at an epoch coincident with chondrule formation. To achieve and maintain such dichotomy, the two chondritic reservoirs had to be kept separated for an extended period of time, over 2-4 million years. Gaps in the disk such as those observed in the simulations presented here create a barrier confining dust on both side of the planet, thus providing an explanation for this dichotomy. Therefore, irrespective of the details of the chondrule formation process, the configuration of the gas flow at the main epoch of chondrule formation was likely akin to that found in our simulations.

	Since vortices are an inevitable consequence of gap formation in our simulations, they were also likely present when chondrules formed. These vortices have different geometries depending on the mass of the planet, the cooling, and the semi-major axis, but they always form inside and outside the gap. The vortices inside are limited in size and are characterised by low temperatures compared to the background temperature. Vortices are also found on both sides of the shock region and accumulate dust grains of varying size \citep{Surville2019}. This leads to enhanced pressure and density of both gas and dust (the magnitude of the enrichment in dust grains of different sizes is under investigation), including potential chondrule precursors in vortices. This would offer a natural explanation for the high quantity of dust that has been converted to chondrules. The pressure increase is also important, because enhanced pressure is required for chondrules to retain the measured concentrations of moderately volatile elements such as Na \citep{Alexander2008}. These elements are otherwise lost during the flash heating of chondrules and are not re-integrated in their minerals during cooling. Vortices could thus be ideal nurseries of chondrules when combined with an additional heat source.

\subsection{ Timing of chondrule formation by bow shocks from Jupiter }
\label{Sect_Discussion_timing}

	Bow shocks of Jupiter can generate chondrule forming conditions with a planet size of at least 0.75 $M_{\jupiter}$ to 1 $M_{\jupiter}$. Ages from chondrules suggest that they formed over a time interval of about 4 Myr with a peak at 2-3 Myr after Calcium-, Aluminium-rich Inclusions (CAIs) \citep{Villeneuve2009, Bollard2017, Pape2019}. This implies that if Jupiter was responsible for the formation of at least part of the chondrule populations, it must have reached a minimum mass of 0.75 $M_{\jupiter}$ before the end of chondrule formation around 4 Myr. Such a timeframe is coherent with astronomical observations implying that planetary disks do not last much longer than 6 Myr, and often less than 3 Myr \citep{HaischJr.2001, Fedele2010}. Moreover, the presence of gas in the disk is essential to generate the nebular shocks to form chondrules. If gas density and pressure are too low, the shocks cannot propagate in the disk preventing chondrules formation by this process.

	This timescale is in agreement with those of some Jupiter growth models \citep{Pollack1996, Lissauer2014}. The processes that governed Jupiter's formation and growth are still debated. The various renditions of the core-accretion scenario published so far assign a formation timescale between 0.5 to 5 Myr to Jupiter, depending on factors such as the opacity of the gaseous envelope and dust grain sedimentation \citep{2014prpl.conf..643H}. In particular, the slow gas accretion phase can last up to a few million years. However, in this phase, the mass of the envelope and the core are comparable and the planet below 0.1 $M_{\jupiter}$. Hence, during this phase, the shock generated by Jupiter are too weak for chondrule formation.

	Recent models \citep{Alibert2018, Kruijer2017} invoke an early formation of Jupiter to explain the isotopic dichotomy of the two classes of chondrites. These models tend to form Jupiter's core early via pebble accretion (within 1 Myr, \cite{Alibert2018}) and the full mass is reached after 4-5 Myr. In these models, the protoplanet grows to 10-20 Earth masses in about 1 Myr. Growth is then slowed down because planetesimal accretion heats the inflating gas envelope. This scenario can explain the dichotomy among chondrites being set up around 1 Myr because the gap generated by the growing Jupiter already limits mixing, at least for larger dust grains such as pebbles \citep{Desch2018}. This creates a sharp structure in the dust distribution. The frequent observations of gaps and rings in the dust of protoplanetary disks \citep{Partnership2015, Andrews2016} supports this scenario. However, while the planet reaches 0.75 $M_{\jupiter}$ at the main epoch of chondrule formation (1 - 4 Myr) in the context of these recent models, it cannot explain the formation of the earlier generations of chondrules ( $< 1$ Myr). Because $M_{\jupiter}$ is insufficient before 1 Myr to generate shocks energetic enough to melt chondrule precursors, another heating mechanism would be required to explain the thermal histories of early chondrules.

	In summary, combining our model results with those of Jupiter's growth and cosmochemical evidence suggests that Jupiter cannot trigger the formation of the entire chondrule population observed in meteorites. A major limiting factor is the growth rate to its final mass, because only a planet with at least 0.75 $M_{\jupiter}$ (see Section \ref{Sect_Results_orbit}) can trigger chondrule formation. It is unlikely that the oldest chondrules in meteorites were formed by nebular shocks of Jupiter, because Jupiter was too small at that time. 

\subsection{ Favourable regions for chondrule formation }
\label{Sect_Discussion_region}

	The subset of the simulations, in which temperatures for chondrule formation are reached, cover a restricted region of the parameter space (Figure \ref{Fig_1}). The results show that ideal conditions are most often met when Jupiter's semi-major axis is between 2.5 and 3 au. Closer locations often heat excessively and vaporise the disk inside Jupiter's orbit, while locations further away are not able to reach sufficiently high temperatures. Interestingly, locations between 2.5 and 3 au were proposed for Jupiter formation based on the position of the water snowline in protoplanetary disk \citep{Ciesla2006, Walsh2011}. Moreover, the evolution model of \cite{Desch2018} also suggests a forming Jupiter at 3 au. This model is based on model calculations of CAI formation and refractory elements distribution. It also proposes the concentration of gas in a pressure bump just outside of Jupiter's orbit as observed in our simulations (Figure \ref{Fig_2}).

	The inferred distances are also compatible with models invoking an inward migration of Jupiter, such as the Grand Tack model \citep{Walsh2011}. In this model, Jupiter forms around 3 au from the Sun and migrates inward, reaching a semi-major axis as close as 1.5 au, before migrating outwards. In this case, chondrules can be formed during the inward migration of a fully grown Jupiter until it reaches 2.5 au. Then, chondrule formation stops because of the very high temperatures reached when the planet is located in the inner disk. Only when the planet "tacks" and migrates outwards, passing through the 2.5 to 3.5 au region, can it form chondrules once more.

\section{ Conclusions }
\label{Sect_Conclusions}

	Our investigation primarily aimed to address the first stage of the chondrule formation process, namely the flash heating stage. Our main conclusions are:
\begin{itemize}
\item The temperature increase in the disk caused by the shocks largely depends on the cooling prescription. Hence, cooling predominately determines whether or not Jupiter-triggered shocks can induce sufficiently high temperatures for chondrule formation. While our simulations show instances in which it is possible to reach temperatures sufficient to melt chondrule precursors in the shocks generated by such a planet, this is unlikely for a MMSN because it can only be achieved under adiabatic or near-adiabatic cooling conditions over the investigated timescales and in regions allowing for subsequent cooling.
\item For all applied cooling models, sufficiently high shock temperatures ($> 1800$ K) are reached with planet orbits interior to 4 au. However, even with the planet in this favourable position, required temperatures for chondrule thermal histories are not achieved when using the most realistic cooling model based on radiative diffusion (see Section \ref{Sect_Results_cooling}). With this "kappa" cooling, such temperatures are only achieved by changing the disk mass to significantly higher values than that of the MMSN (see Sections \ref{Sect_Results_cooling}, \ref{Sect_Discussion}). Simulations with a more massive nebula suggest that shocks from massive planets have the potential to form chondrules in different settings, such as exoplanetary systems. In such massive disks, however, it is still unclear if rapid cooling following shock heating can be achieved locally, which is also required for chondrule formation. 
\item Large vortices forming in the disk under the planet's influence increase dust density and thus can concentrate dust acting as chondrule precursors. They also create regions of high gas pressure and density that are favourable for retaining moderately volatile elements in the chondrule formation environment. As a result, these vortices are locations where a large quantity of dust could experience conditions favourable to chondrule formation.
\end{itemize}

	Other considerations are also coherent with nebular shocks from Jupiter as chondrule forming scenario. The presence of a massive planet carves a gap in the disk in which density and pressure of the gas are highly reduced. Such a gap limits exchange of material between the inner and outer part of the disk and is an explanation for the isotopic dichotomy reported between meteorites that formed in the inner (enstatite and ordinary chondrites) and the outer disk (carbonaceous chondrites). 
The growth times for Jupiter estimated from early solar system models are coherent with chondrule formation by shocks from Jupiter, but only for the late generations of chondrules.

\begin{acknowledgements}

	This work has been carried out within the frame of the National Center for Competence in Research {\it{PlanetS}} supported by the Swiss National Science Foundation (SNSF). The authors acknowledge the financial support of the SNSF. Numerical simulations were performed on the Euler cluster of the ETHZ.

\end{acknowledgements}

\bibliographystyle{aa}
\bibliography{Biblio}

\begin{thebibliography}{53}
\expandafter\ifx\csname natexlab\endcsname\relax\def\natexlab#1{#1}\fi

\bibitem[{Alexander {et~al.}(2008)Alexander, Grossman, Ebel, \&
  Ciesla}]{Alexander2008}
Alexander, C. M.~O., Grossman, J.~N., Ebel, D.~S., \& Ciesla, F.~J. 2008,
  Science (New York, N.Y.), 320, 1617

\bibitem[{Alibert {et~al.}(2018)Alibert, Venturini, Helled, Ataiee, Burn,
  Senecal, Benz, Mayer, Mordasini, Quanz, \&
  Sch{\"{o}}nb{\"{a}}chler}]{Alibert2018}
Alibert, Y., Venturini, J., Helled, R., {et~al.} 2018, Nature Astronomy, 2, 873

\bibitem[{Andrews {et~al.}(2016)Andrews, Wilner, Zhu, Birnstiel, Carpenter,
  Perez, Bai, Oberg, Hughes, Isella, \& Ricci}]{Andrews2016}
Andrews, S.~M., Wilner, D.~J., Zhu, Z., {et~al.} 2016, The Astrophysical
  Journal Letters, 820, 0

\bibitem[{Bell \& Lin(1994)}]{Bell1994}
Bell, K.~R. \& Lin, D. N.~C. 1994, The Astrophysical Journal, 427, 987

\bibitem[{Bollard {et~al.}(2017)Bollard, Connelly, Whitehouse, Pringle, Bonal,
  J{\o}rgensen, Nordlund, Moynier, \& Bizzarro}]{Bollard2017}
Bollard, J., Connelly, J.~N., Whitehouse, M.~J., {et~al.} 2017, Science
  Advances, 3, e1700407

\bibitem[{Ciesla \& Cuzzi(2006)}]{Ciesla2006}
Ciesla, F.~J. \& Cuzzi, J.~N. 2006, Icarus, 181, 178

\bibitem[{de~Val-Borro {et~al.}(2007)de~Val-Borro, Artymowicz, Angelo, \&
  Peplinski}]{DeVal-Borro2007}
de~Val-Borro, M., Artymowicz, P., Angelo, G.~D., \& Peplinski, A. 2007,
  Astronomy {\&} Astrophysics, 471, 1043

\bibitem[{Desch \& Connolly(2002)}]{Desch2002}
Desch, S.~J. \& Connolly, H. C.~J. 2002, Meteoritics {\&} Planetary Science,
  37, 183

\bibitem[{Desch {et~al.}(2018)Desch, Kalyaan, \& Alexander}]{Desch2018}
Desch, S.~J., Kalyaan, A., \& Alexander, C. M.~O. 2018, The Astrophysical
  Journal Supplement Series, 238, 11

\bibitem[{Dullemond {et~al.}(2016)Dullemond, Harsono, Stammler, \&
  Johansen}]{Dullemond2016a}
Dullemond, C.~P., Harsono, D., Stammler, S.~M., \& Johansen, A. 2016, The
  Astrophysical Journal, 832, 91

\bibitem[{Dullemond {et~al.}(2014)Dullemond, Stammler, \&
  Johansen}]{Dullemond2014}
Dullemond, C.~P., Stammler, S.~M., \& Johansen, A. 2014, The Astrophysical
  Journal, 794, 91

\bibitem[{Fedele {et~al.}(2010)Fedele, Ancker, Henning, Jayawardhana, \&
  Oliveira}]{Fedele2010}
Fedele, D., Ancker, M. E. V.~D., Henning, T., Jayawardhana, R., \& Oliveira,
  J.~M. 2010, Astronomy {\&} Astrophysics, 510, A72

\bibitem[{Fu {et~al.}(2014)Fu, Li, Lubow, Li, \& Liang}]{Fu2014}
Fu, W., Li, H., Lubow, S., Li, S., \& Liang, E. 2014, The Astrophysical
  Journal, 795, L39

\bibitem[{Gong {et~al.}(2019)Gong, Zheng, Lin, Silsbee, Baruteau, \&
  Mao}]{Gong2019}
Gong, M., Zheng, X., Lin, D. N.~C., {et~al.} 2019, The Astrophysical Journal,
  883, 164

\bibitem[{Guillot {et~al.}(1997)Guillot, Gautier, \& Hubbard}]{Guillot1997}
Guillot, T., Gautier, D., \& Hubbard, W.~B. 1997, Icarus, 539, 534

\bibitem[{{Haisch, Jr.} {et~al.}(2001){Haisch, Jr.}, Lada, \&
  Lada}]{HaischJr.2001}
{Haisch, Jr.}, K.~E., Lada, E.~A., \& Lada, C.~J. 2001, The Astrophysical
  Journal, 553, L153

\bibitem[{Hayashi(1981)}]{Hayashi1981}
Hayashi, C. 1981, Supplement of the Progress of Theoretical Physics, 70, 35

\bibitem[{Helled {et~al.}(2014)Helled, Bodenheimer, Podolak, Boley, Meru,
  Nayakshin, Fortney, Mayer, Alibert, \& Boss}]{2014prpl.conf..643H}
Helled, R., Bodenheimer, P., Podolak, M., {et~al.} 2014, in Protostars and
  Planets VI, ed. H.~Beuther, R.~S. Klessen, C.~P. Dullemond, \& T.~Henning,
  643

\bibitem[{Hewins(1988)}]{1988mess.book..660H}
Hewins, R. 1988, {Experimental studies of chondrules.}, ed. J.~F. Kerridge \&
  M.~S. Matthews, 660--679

\bibitem[{Hewins {et~al.}(2005)Hewins, Connolly, \& Libourel}]{Hewins2005}
Hewins, R.~H., Connolly, H. C.~J., \& Libourel, G. 2005, Chondrites and the
  Protoplanetary Disk, 341, 286

\bibitem[{Hor{\'{a}}nyi {et~al.}(1995)Hor{\'{a}}nyi, Morfill, Goertz, \&
  Levy}]{Horanyi1995}
Hor{\'{a}}nyi, M., Morfill, G., Goertz, C.~K., \& Levy, E.~H. 1995, Icarus,
  114, 174

\bibitem[{Iida {et~al.}(2001)Iida, Nakamoto, \& Susa}]{Iida2001}
Iida, A., Nakamoto, T., \& Susa, H. 2001, Icarus, 153, 430

\bibitem[{Jones(2012)}]{Jones2012}
Jones, R.~H. 2012, Meteoritics and Planetary Science, 47, 1176

\bibitem[{Kruijer {et~al.}(2017)Kruijer, Burkhardt, Budde, \&
  Kleine}]{Kruijer2017}
Kruijer, T.~S., Burkhardt, C., Budde, G., \& Kleine, T. 2017, Proceedings of
  the National Academy of Sciences, 114, 6712

\bibitem[{Les \& Lin(2015)}]{Les2015}
Les, R. \& Lin, M.-k. 2015, Monthly Notices of the Royal Astronomical Society,
  150, 1503

\bibitem[{Leya {et~al.}(2008)Leya, Sch{\"{o}}nb{\"{a}}chler, \&
  Wiechert}]{Leya2008}
Leya, I., Sch{\"{o}}nb{\"{a}}chler, M., \& Wiechert, U. 2008, Earth and
  Planetary Science Letters, 266, 233

\bibitem[{Lichtenberg {et~al.}(2018)Lichtenberg, Golabek, Dullemond,
  Sch{\"{o}}nb{\"{a}}chler, Gerya, \& Meyer}]{Lichtenberg2018}
Lichtenberg, T., Golabek, G.~J., Dullemond, C.~P., {et~al.} 2018, Icarus, 302,
  27

\bibitem[{Lissauer(1987)}]{Lissauer1987}
Lissauer, J.~J. 1987, Icarus, 69, 249

\bibitem[{Lissauer(1993)}]{Lissaue1993}
Lissauer, J.~J. 1993, Annual Review of Astronomy and Astrophysics, 31, 129

\bibitem[{Lissauer {et~al.}(2014)Lissauer, Marcy, Bryson, Rowe, Jontof-hutter,
  Agol, Borucki, Carter, Ford, Gilliland, Kolbl, Star, Steffen, \&
  Torres}]{Lissauer2014}
Lissauer, J.~J., Marcy, G.~W., Bryson, S.~T., {et~al.} 2014, The Astrophysical
  Journal, 784, 44

\bibitem[{Liu {et~al.}(2018)Liu, Dipierro, Ragusa, Lodato, Herczeg, Long,
  Boehler, Menard, Johnstone, Pascucci, Pinilla, Plas, Cabrit, Fischer,
  Hendler, \& Manara}]{Liu2018}
Liu, Y., Dipierro, G., Ragusa, E., {et~al.} 2018, arXiv

\bibitem[{Lofgren(1996)}]{1996cpd..conf..187L}
Lofgren, G. 1996, in Chondrules and the Protoplanetary Disk, 187--196

\bibitem[{Malik {et~al.}(2015)Malik, Meru, Mayer, \& Meyer}]{Malik2015}
Malik, M., Meru, F., Mayer, L., \& Meyer, M. 2015, The Astrophysical Journal,
  802, 56

\bibitem[{Mercer {et~al.}(2018)Mercer, Stamatellos, \& Dunhill}]{Mercer2018}
Mercer, A., Stamatellos, D., \& Dunhill, A. 2018, Monthly Notices of the Royal
  Astronomical Society, 478, 3478

\bibitem[{Mohanty {et~al.}(2013)Mohanty, Greaves, Mortlock, Pascucci, Scholz,
  Thompson, Apai, Lodato, \& Looper}]{Mohanty2013}
Mohanty, S., Greaves, J., Mortlock, D., {et~al.} 2013, The Astrophysical
  Journal, 773, 168

\bibitem[{Morris \& Desch(2010)}]{Morris2010}
Morris, M.~A. \& Desch, S.~J. 2010, The Astrophysical Journal, 722, 1474

\bibitem[{Morris {et~al.}(2016)Morris, Weidenschilling, \& Desch}]{Morris2016}
Morris, M.~A., Weidenschilling, S.~J., \& Desch, S.~J. 2016, Meteoritics and
  Planetary Science, 51, 870

\bibitem[{Pape {et~al.}(2019)Pape, Mezger, Bouvier, \& Baumgartner}]{Pape2019}
Pape, J., Mezger, K., Bouvier, A.~S., \& Baumgartner, L.~P. 2019, Geochimica et
  Cosmochimica Acta, 244, 416

\bibitem[{Partnership {et~al.}(2015)Partnership, Fomalont, Vlahakis, Corder,
  Remijan, Barkats, \& Lucas}]{Partnership2015}
Partnership, A., Fomalont, E.~B., Vlahakis, C., {et~al.} 2015, The
  Astrophysical Journal Letters, 1, 1

\bibitem[{Pollack {et~al.}(1996)Pollack, Hubickyj, Bodenheimer, Lissauer,
  Podolak, \& Greenzweig}]{Pollack1996}
Pollack, J.~B., Hubickyj, O., Bodenheimer, P., {et~al.} 1996, Icarus, 124, 62

\bibitem[{Rafikov(2002)}]{Rafikov2002}
Rafikov, R.~R. 2002, The Astrophysical Journal, 569, 997

\bibitem[{Richert {et~al.}(2015)Richert, Lyra, Boley, Low, \&
  Turner}]{Richert2015}
Richert, A. J.~W., Lyra, W., Boley, A., Low, M.-m.~M., \& Turner, N. 2015, The
  Astrophysical Journal, 804, 1

\bibitem[{Safronov(1991)}]{Safronov1991}
Safronov, V.~S. 1991, Icarus, 94, 260

\bibitem[{Sanders \& Scott(2012)}]{Sanders2012}
Sanders, I.~S. \& Scott, E. R.~D. 2012, Meteoritics and Planetary Science, 47,
  2170

\bibitem[{Stamatellos {et~al.}(2007)Stamatellos, Whitworth, Bisbas, \&
  Goodwin}]{Stamatellos2007}
Stamatellos, D., Whitworth, A.~P., Bisbas, T., \& Goodwin, S. 2007, Astronomy
  {\&} Astrophysics, 475, 37

\bibitem[{Stammler \& Dullemond(2014)}]{Stammler2014}
Stammler, S.~M. \& Dullemond, C.~P. 2014, Icarus, 242, 1

\bibitem[{Surville \& Barge(2015)}]{Surville2015}
Surville, C. \& Barge, P. 2015, Astronomy {\&} Astrophysics, 579, A100

\bibitem[{Surville \& Mayer(2019)}]{Surville2019}
Surville, C. \& Mayer, L. 2019, The Astrophysical Journal, 883, 176

\bibitem[{Surville {et~al.}(2016)Surville, Mayer, \& Lin}]{Surville2016}
Surville, C., Mayer, L., \& Lin, D. N.~C. 2016, The Astrophysical Journal, 831,
  82

\bibitem[{Trinquier {et~al.}(2007)Trinquier, Birck, \&
  All{\`{e}}gre}]{Trinquier2007}
Trinquier, A., Birck, J.-L., \& All{\`{e}}gre, C.~J. 2007, The Astrophysical
  Journal, 655, 1179

\bibitem[{Villeneuve {et~al.}(2009)Villeneuve, Chaussidon, \&
  Libourel}]{Villeneuve2009}
Villeneuve, J., Chaussidon, M., \& Libourel, G. 2009, Science, 325, 985

\bibitem[{Walsh {et~al.}(2011)Walsh, Morbidelli, Raymond, O'Brien, \&
  Mandell}]{Walsh2011}
Walsh, K.~J., Morbidelli, A., Raymond, S.~N., O'Brien, D.~P., \& Mandell, A.~M.
  2011, Nature, 475, 206

\bibitem[{Zhu {et~al.}(2014)Zhu, Stone, Rafikov, \& Bai}]{Zhu2014}
Zhu, Z., Stone, J.~M., Rafikov, R.~R., \& Bai, X.-n. 2014, The Astrophysical
  Journal, 785, 122

\end{thebibliography}

\end{document}